\documentclass[11pt]{article}
\usepackage{moriond,epsfig}

\def\be{\begin{equation}}
\def\ee{\end{equation}}
\def\bea{\begin{eqnarray}}
\def\eea{\end{eqnarray}}

\begin{document}
\title{SHOT NOISE OF WEAK COTUNNELING CURRENT:\\
NON-EQUILIBRIUM FLUCTUATION-DISSIPATION THEOREM}

\author{\underline{EUGENE V. SUKHORUKOV},
DANIEL LOSS}

\address{University of Basel, Basel, Switzerland}

\maketitle\abstracts{ We study the noise of the cotunneling
current through one or several tunnel-coupled quantum dots in the
Coulomb blockade regime. We consider the regime of weak 
(elastic and inelastic) cotunneling, and prove a non-equilibrium
fluctuation-dissipation theorem which leads to a universal
expression for the noise-to-current ratio (Fano factor). }

\section{Introduction}
\label{introduction}

In recent years, there has been great interest in
the shot noise 
in mesoscopic
systems,~\cite{review} because it contains additional information 
about correlations, which is not contained, e.g., in
the linear response conductance. The shot noise is characterized 
by the Fano factor $F=S/eI$, the dimensionless ratio of the zero-frequency
noise power $S$ to the average current $I$.
While it assumes the Poissonian value $F=1$ in the absence of correlations,
it becomes suppressed or enhanced when correlations set in as e.g.
imposed by the Pauli principle or due to interaction effects.
In the present paper we study the shot noise of the 
cotunneling~\cite{averinazarov,Glattli} current.
We consider the transport through a quantum-dot system (QDS)
in the Coulomb blockade (CB) regime, in which the quantization of
charge on the QDS leads to a suppression of the sequential tunneling
current except under certain resonant conditions.  We consider the transport
away from these resonances and study the next-order contribution to
the current~\footnote{
The majority of papers on the noise of quantum dots consider
the sequential tunneling regime, where a classical description
(``orthodox'' theory) is applicable.~\cite{AL}
In this regime the noise is generally suppressed below its 
full Poissonian value $F=1$.
This suppression can be interpreted~\cite{SBL} 
as being a result of the natural correlations imposed 
by charge conservation.} (see Fig.~1). 
One might expect that the cotunneling,
being a two-particle process, leads to strong correlations in the shot
noise and to the deviation of the Fano factor
from its Poissonian value $F=1$. 
As it has been found recently,~\cite{SBL} 
this is indeed the case for the regime of strong cotunneling, 
i.e. when the cotunneling rate $I/e$ is large compared
to the intrinsic relaxation rate $w_{\rm in}$ of the QDS to its equilibrium
state due to the coupling to the environment, $I/e\gg w_{\rm in}$.
However we find here, that in the weak cotunneling 
regime, $I/e\ll w_{\rm in}$, the zero-frequency noise  takes on its
Poissonian value, as first obtained for a
special case.~\cite{LS} This result is
generalized here, and we find a universal relation
between noise and current for the QDS
in the first nonvanishing order in the tunneling
perturbation. Because of the universal
character of this result (Eq.~\ref{DB-FDT})
we call it the nonequilibrium 
fluctuation-dissipation theorem (FDT)~\cite{Rogovin}
in analogy with linear response theory.

\section{Model system}

In general, the QDS can contain several dots, which can be
coupled by tunnel junctions, the double dot (DD) being a particular
example.~\cite{LS}
The QDS is assumed to be weakly coupled to external metallic leads which are
kept at equilibrium with their associated reservoirs at the chemical
potentials $\mu_l$, $l=1,2$,  where the  currents $I_l$ can be measured
and the average current $I$
through the QDS is defined by Eq.~\ref{current-noise}.
Using a standard tunneling Hamiltonian approach,~\cite{Mahan}
we write
\begin{eqnarray}
&& H=H_0+V\,,\quad H_0=H_L+H_S+H_{\rm int}\,, \label{Hamiltonian}   \\
&& H_L=\sum_{l=1,2}\sum_k\varepsilon_{k}c_{lk}^{\dag}c_{lk}\,, \quad
H_S=\sum_p\varepsilon_pd_p^{\dag}d_p\,, \label{QDS-leads} \\
&& V=\sum_{l=1,2}(D_l+D^{\dag}_l),\quad
D_l=\sum_{k,p}T_{lkp}c_{lk}^{\dag}d_p\,,\label{tunneling}
\end{eqnarray}
where  the terms $H_L$ and $H_S$  describe
the leads and QDS, respectively (with $k$ and $p$
from a complete set of quantum numbers),and
tunneling between leads  and  QDS is described
by the perturbation $V$. The interaction term $H_{\rm int}$
does not need to be specified for our proof of the universality 
of noise.
The $N$-electron QDS is in the cotunneling regime where there is
a finite energy cost
$\Delta_{\pm}(l,N)>0$
for the electron tunneling from the Fermi level of the lead $l$
to the QDS ($+$) and vice versa ($-$),
so that only processes of second order in $V$ are allowed.

\begin{figure}[t]
\begin{minipage}{75mm}
\epsfxsize=75mm
\epsfbox{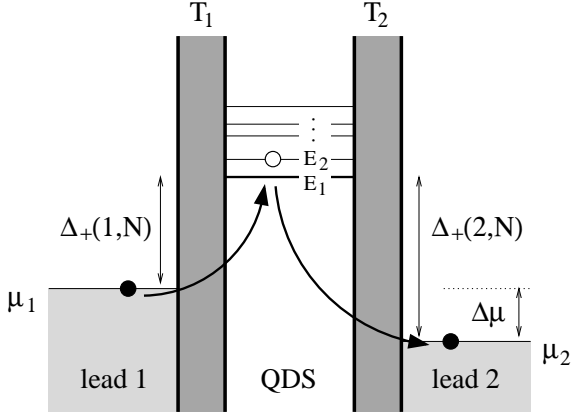}
\end{minipage}\hfill
\parbox{80mm}{\caption{
The quantum dot system (QDS) is coupled to
two external leads $l=1,2$ via tunneling barriers.
The tunneling between the QDS and the leads
is parametrized by the tunneling amplitudes $T_l$,
where the lead and QDS quantum numbers $k$ and $p$ have
been dropped for simplicity.
The leads are at the chemical potentials $\mu_{1,2}$, with an applied
bias $\Delta\mu=\mu_1 -\mu_2$.
The eigenstates of the QDS with one
added electron ($N+1$ electrons in total) are indicated by their energies
$E_1,E_2,\ldots$.
In the cotunneling regime there is
a finite energy cost
$\Delta_{\pm}(l,N)>0$
for the electron tunneling from the Fermi level of the lead $l$
to the QDS ($+$) and vice versa ($-$),
so that only processes of second order in $V$ (visualized by two arrows) 
are allowed.}}
\end{figure}

To describe the transport through the QDS we apply standard methods~\cite{Mahan}
and adiabatically switch on the perturbation $V$ in the distant past,
$t=t_0\to - \infty$. The perturbed state of the system
is described by the time-dependent density matrix
$\rho(t)=e^{-iH(t-t_0)}\rho_0 e^{iH(t-t_0)}$,
with $\rho_0$ being the grand canonical density matrix of the
unperturbed system, $\rho_0=Z^{-1}e^{-K/k_BT}$,
where we set $K=H_0-\sum_l\mu_lN_l$.
Because of tunneling the total number of electrons in each lead
$N_l=\sum_kc_{lk}^{\dag}c_{lk}$ is no longer conserved.
For the outgoing currents $\hat I_l=e\dot N_l$ we have
\begin{equation}
\hat I_l=ei\left[V,N_l\right] =ei(D^{\dag}_l-D_l)\,.
\label{currents}
\end{equation}
The observables of interest are the average current $I\equiv I_2=-I_1$
through the QDS,
and the spectral density of the noise
$S_{ll'}(\omega)=\int dt S_{ll'}(t)\exp(i\omega t)$,
\begin{equation}
I_l={\rm Tr}\rho(0)\hat I_l,\quad
S_{ll'}(t)={\rm Re}\,{\rm Tr}\,\rho(0)\delta I_l(t)\delta I_{l'}(0)\,,
\label{current-noise}
\end{equation}
where $\delta I_l=\hat I_l-I_l$.
Below we will use the interaction representation
where Eq.~\ref{current-noise} can be rewritten by replacing
$\rho(0)\to\rho_0$ and $\hat I_l(t)\to U^{\dag}(t)\hat I_l(t)U(t)$, with
\begin{equation}
U(t)=T\exp\left[-i\int^{t}_{-\infty}dt'\,V(t')\right]\,.
\label{U-Operator}
\end{equation}
In this representation, the time dependence of all operators is governed by the
unperturbed Hamiltonian $H_0$.

\section{Non-equilibrium fluctuation-dissipation theorem}
\label{FDT-Double}

We note that the two currents $\hat I_l$ are not independent,
because $[\hat I_1,\hat I_2]\neq 0$, and thus all
correlators $S_{ll'}$ are nontrivial.
The charge accumulation on the QDS for a time of order
$\Delta_{\pm}^{-1}$ leads to an additional contribution to the noise
at finite frequency $\omega$. Thus, we expect
that for $\omega\sim\Delta_{\pm}$ the correlators
$S_{ll'}$ cannot be expressed through the steady-state
current $I$ only and thus $I$ has to be complemented by some
other dissipative counterparts, such as differential conductances
$G_{ll'}$. On the other hand, at low enough frequency, 
$\omega \ll \Delta_{\pm}$,
the charge conservation on the QDS requires
$\delta I_s=(\delta I_2+\delta I_1)/2\approx 0$.
Below we concentrate on the limit of low frequency
and neglect contributions of order of $\omega/\Delta_{\pm}$ to the noise power.
In  the Appendix we prove that $S_{ss}\sim (\omega/\Delta_{\pm})^2$ 
(see Eq.~\ref{A09}),
and this allows us to redefine the current and the noise power as
$I\equiv I_d=(I_2-I_1)/2$ and 
$S(\omega)\equiv S_{dd}(\omega)$.~\footnote{
We note that charge fluctuations,
$\delta Q(t)\!=\!2\!\int_{-\infty}^{t}\!dt' \delta I_s(t')$,
on a QDS are also relevant for device
applications such as SET.~\cite{problem}
While we focus on current fluctuations in the present paper, we mention here
that in the cotunneling regime
the noise power $\langle \delta Q^2\rangle_{\omega}$
does not vanish at zero frequency,
$\langle \delta Q^2\rangle_{\omega=0}=
4\omega^{-2}S_{ss}(\omega)|_{\omega\to 0}\neq 0$.
Our formalism is also suitable for studying such charge fluctuations;
this will be addressed elsewhere.}
In addition we require that the QDS is in the cotunneling regime,
i.e. the temperature is low enough, $k_BT\ll \Delta_{\pm}$, although
the bias $\Delta\mu$ is arbitrary as soon as the sequential 
tunneling to the dot is forbidden, $\Delta_{\pm}>0$.
In this limit the current through a QDS arises due to the direct
hopping of an electron from one lead to another (through a virtual state
on the dot) with an amplitude which depends on the energy cost $\Delta_{\pm}$
of a virtual state. Although this process can change the state
of the QDS (inelastic cotunneling), the fast energy relaxation
in the weak cotunneling regime, $w_{\rm in}\gg I/e$, immediately
returns it to the equilibrium state (for the opposite case,
see Ref.~\cite{SBL}). This allows
us to apply a perturbation expansion with respect to tunneling $V$
and to keep only first nonvanishing contributions, which we do next.

It is convenient to introduce the notation
$\bar D_l(t)\equiv\int_{-\infty}^{t}dt'\, D_l(t')$.
We notice that all relevant matrix elements,
$\langle N| D_l(t)|N+1\rangle\sim e^{-i\Delta_{+}t}$,
$\langle N-1| D_l(t)|N\rangle\sim e^{i\Delta_{-}t}$,
are fast oscillating functions of time. Thus, under the above
conditions we can write $\bar D_l(\infty)=0$,
and even more general, $\int_{-\infty}^{+\infty}dt\, D_l(t)e^{\pm i\omega t}=0$
(note that we have assumed earlier that $\omega\ll \Delta_\pm$).
Using these equalities and the cyclic property
of the trace we obtain the following results (for details of the derivation,
see Appendix),
\begin{eqnarray}
&&I=e\int\limits_{-\infty}^{\infty}dt\,
\langle [A^{\dag}(t),A(0)]\rangle,\qquad
A=D_2\bar D^{\dag}_1+D^{\dag}_1\bar D_2\,,
\label{DB-current}\\
&&S(\omega)=e^2\int\limits^{\infty}_{-\infty}dt\,
\cos (\omega t)\langle\{A^{\dag}(t),A(0)\}\rangle\,,
\label{DB-noise}
\end{eqnarray}
where we have dropped a small contribution of order
$\omega/\Delta_{\pm}$ and used the notation 
$\langle\ldots\rangle={\rm Tr}\rho_0(\ldots)$.

Next we apply the spectral decomposition
to the correlators Eqs.~\ref{DB-current} and~\ref{DB-noise},
a similar procedure to that which also leads
to the equilibrium fluctuation-dissipation theorem.
The crucial observation is that $[H_0,N_l]=0$, $l=1,2$.
Therefore, we are allowed to use for our spectral decomposition the basis
$|{\bf n}\rangle=| E_{{\bf n}},N_1,N_2\rangle$ of eigenstates
of the operator $K=H_0-\sum_l\mu_l N_l$, which also diagonalizes the grand-canonical
density matrix $\rho_0$, $\rho_{{\bf n}}=\langle {\bf n}|\rho_0 |{\bf n} \rangle
=Z^{-1}\exp[-E_{{\bf n}}/k_BT]$.
We introduce the spectral function,
\begin{equation}
{\cal A}(\omega) =
2\pi \sum_{ {\bf n}, {\bf m}}(\rho_{{\bf n}} +\rho_{{\bf m}})
|\langle {\bf m}|A|{\bf n}\rangle|^2
\delta(\omega+E_{{\bf n}}-E_{{\bf m}})\,,
\label{spectral}
\end{equation}
and rewrite Eqs.~\ref{DB-current} and~\ref{DB-noise}
in the matrix form in the basis $| {\bf n}\rangle$
taking into account that
the operator $A$, which plays the role of the effective cotunneling amplitude, 
creates (annihilates) an electron in the lead 2 (1) 
(see Eqs.~\ref{tunneling} and~\ref{DB-current}). 
We obtain following expressions
\begin{eqnarray}
&&  I(\Delta\mu)
=e\tanh\left[\frac{\Delta\mu}{2k_BT}\right]{\cal A}(\Delta\mu)\,,
\label{DB-current2} \\
&& S(\omega,\Delta\mu)=\frac{e^2}{2}\sum_{\pm}
{\cal A}(\Delta\mu\pm\omega)\,.
\label{DB-noise2}
\end{eqnarray}
We note that because of additional integration over time $t$ in the 
amplitude $A$ (see Eq.~\ref{DB-current}), the spectral density 
${\cal A}$ depends on $\mu_1$ and $\mu_2$ separately.
However, away from the resonances, $\omega\ll\Delta_\pm$,
only $\Delta\mu$-dependence is essential, and thus ${\cal A}$
can be regarded as being one-parameter 
function.~\footnote{To be more precise, we neglect small
$\omega$-shift of the energy denominators $\Delta_\pm$, which is equivalent
to neglecting small terms of order $\omega/\Delta_\pm$ in Eq.~\ref{DB-noise2}.}
Comparing Eqs.~\ref{DB-current2} and~\ref{DB-noise2},
we obtain
\begin{equation}
 S(\omega,\Delta\mu)=\frac{e}{2}\sum_{\pm}
\coth\left[\frac{\Delta\mu\pm\omega}{2k_BT}\right]
I(\Delta\mu\pm\omega)
\label{DB-FDT}
\end{equation}
up to small terms on the order of $\omega/\Delta_\pm$. 
This equation represents our nonequilibrium FDT for the
transport through a QDS in the weak cotunneling regime.
A special case with $T, \omega=0$, giving $S=eI$, has
been derived earlier.~\cite{LS}
To conclude this section we would like to list again the conditions
used in the derivation.
The universality of noise to current relation
Eq.~\ref{DB-FDT} proven here is valid in the regime in which
it is sufficient to keep the  first nonvanishing order in the tunneling $V$
which contributes to transport and noise.
This means that  the QDS is in the weak cotunneling regime
with $\omega, k_BT \ll \Delta_{\pm }$, and
$I/e\ll w_{\rm in}$.

\section*{Acknowledgements}
This work was supported by the Swiss NSF and by the DARPA
program SPINS (DAAD19-01-0327).

\section*{Appendix}

In this Appendix we present the derivation of Eqs.~\ref{DB-current}
and \ref{DB-noise}.
In order to simplify the intermediate steps, we use the
notation $\bar O(t)\equiv\int_{-\infty}^{t}dt'O(t')$ for any operator $O$,
and $O(0)\equiv O$.
We notice that, if an operator $O$ is a linear function of operators $D_l$ and $D_l^{\dag}$,
then $\bar O(\infty)=0$ (see the discussion in Sec.~\ref{FDT-Double}).
Next, the currents can be represented as the difference and the sum of $\hat I_1$
and $\hat I_2$,
\begin{eqnarray}
\hat I_d &=& (\hat I_2-\hat I_1)/2=ie(X^{\dag}-X)/2\,,
\label{A01}\\
\hat I_s &=& (\hat I_1+\hat I_2)/2=ie(Y^{\dag}-Y)/2\,,
\label{A02}
\end{eqnarray}
where $X=D_2+D_1^{\dag}$, and $Y=D_1+D_2$.
While for the perturbation we have
\begin{equation}
V=X+X^{\dag}=Y+Y^{\dag}\,.
\label{A03}
\end{equation}
First we concentrate on the derivation
of Eq.~\ref{DB-current} and redefine the average current 
Eq.~\ref{current-noise} as
$I=I_d$ (which gives the same result anyway, because the average number
of electrons on the QDS does not change $I_s =0$).

To proceed with our derivation, we make use of Eq.~\ref{U-Operator} and expand
the current up to fourth order in $T_{lkp}$:
\begin{equation}
I = i\int\limits^{0}_{-\infty}dt\int\limits^{t}_{-\infty}dt'
\langle \hat I_dV(t)V(t')\bar V(t')\rangle 
- i\int\limits^{0}_{-\infty}dt \langle \bar V\hat I_dV(t)\bar V(t)\rangle
+ {\rm c.c.}\,
\label{A04}
\end{equation}
Next, we use the cyclic property of trace to shift the time dependence to $\hat I_d$.
Then we complete the integral over time $t$ and use $\bar I_d(\infty)=0$. This procedure
allows us to combine first and second term in Eq.~\ref{A04},
\begin{equation}
I=-i\int\limits^{0}_{-\infty}dt
\langle [\bar I_dV+\bar V\hat I_d]V(t)\bar V(t)\rangle+{\rm c.c.}
\label{A05}\,
\end{equation}
Now, using Eqs.~\ref{A01} and~\ref{A03}
we replace operators in Eq.~\ref{A05} with $X$ and $X^{\dag}$ in two steps:
$I=e\int^{0}_{-\infty}dt
\langle [\bar X^{\dag}X^{\dag}-\bar XX]V(t)\bar V(t)\rangle+{\rm c.c.}$,
where some terms cancel exactly. Then we work with $V(t)\bar V(t)$ and
notice that some terms cancel, because they are linear in $c_{lk}$ and
$c_{lk}^{\dag}$. Thus we obtain
$I = e\int^{0}_{-\infty}dt
\langle [\bar X^{\dag}X^{\dag}-\bar XX]
[X^{\dag}(t)\bar X^{\dag}(t)+X(t)\bar X(t)]\rangle
+{\rm c.c.}$.
Two terms $\bar XXX\bar X$ and
$\bar X^{\dag}X^{\dag}X^{\dag}\bar X^{\dag}$ describe tunneling
of two electrons from the same lead, and therefore they do not contribute
to the normal current. We then combine all other terms to extend the integral to $+\infty$,
\begin{equation}
I = e\int\limits^{\infty}_{-\infty}dt
\langle\bar X^{\dag}(t)X^{\dag}(t)X\bar X
-\bar XXX^{\dag}(t)\bar X^{\dag}(t)\rangle\,
\label{A07}
\end{equation}
Finally, we use
$\int^{\infty}_{-\infty}dt X(t)\bar X(t)=-\int^{\infty}_{-\infty}dt \bar X(t)X(t)$
(since $\bar X(\infty)=0$) to get Eq.~\ref{DB-current} with $A=X\bar X$.
Here, again, we drop
terms $D^{\dag}_1\bar D^{\dag}_1$ and $D_2\bar D_2$ responsible for
tunneling of two electrons from the same lead, and obtain $A$
as in Eq.~\ref{DB-current}.

Next, we derive Eq.~\ref{DB-noise} for the noise power.
At small frequencies $\omega\ll \Delta_{\pm}$ fluctuations
of $I_s$ are suppressed because of charge conservation (see below),
and we can replace $\hat I_2$ in the correlator Eq.~\ref{current-noise} 
with $\hat I_d$.
We expand $S({\omega})$ up to fourth order in $T_{lkp}$,
use $\int_{-\infty}^{+\infty}dt\, \hat I_d(t)e^{\pm i\omega t}=0$,
and repeat the steps leading to Eq.~\ref{A05}. Doing this we
obtain,
\begin{equation}
S(\omega)=-\int\limits^{\infty}_{-\infty}dt \cos(\omega t)
\langle [\bar V(t),\hat I_d(t)][\bar V, \hat I_d]\rangle\,.
\label{A08}
\end{equation}
Then, we replace $V$ and $\hat I_d$ with $X$ and $X^{\dag}$.
We again keep only terms relevant
for cotunneling,
and in addition we neglect terms of order $\omega/\Delta_{\pm}$
(applying same arguments as before, see Eq.~\ref{A09}).
We then arrive at Eq.~\ref{DB-noise}
with the operator $A$ given by Eq.~\ref{DB-current}.

Finally, in order to show that fluctuations of $I_s$ are suppressed,
we replace $\hat I_d$ in Eq.~\ref{A08} with $\hat I_s$,
and then use the operators
$Y$ and $Y^{\dag}$ instead of $X$ and $X^{\dag}$.
In contrast to Eq.~\ref{A07} terms such as
$\bar Y^{\dag}Y^{\dag}Y\bar Y$ do not contribute,
because they contain integrals  of the form
$\int^{\infty}_{-\infty}dt\cos(\omega t) D_{l}(t)\bar D_{l'}(t)=0$.
The only nonzero contribution can be written as
\begin{equation}
S_{ss}(\omega)=\frac{e^2\omega^2}{4}\int\limits^{\infty}_{-\infty}dt \cos(\omega t)
\langle [\bar Y^{\dag}(t),\bar Y(t)][\bar Y^{\dag},\bar Y]\rangle\,,
\label{A09}
\end{equation}
where we have used integration by parts and the property $\bar Y(\infty)=0$.
Compared to Eq.~\ref{DB-noise} this expression contains an additional
integration over $t$, and thereby it is of order
$(\omega/\Delta_{\pm})^2$.

\end{document}